# Multi-class versus One-class classifier in spontaneous speech analysis oriented to Alzheimer Disease diagnosis

K. López-de-Ipiña[1], Marcos Faundez-Zanuy[2], Jordi Sole[3], Fernando Zelarin[5], Pilar Calvo[1]

[1]Universidad del País Vasco/Euskal Herriko Unibertsitatea, Europa Pz 1, 20008 Donostia Spain
`karmele.ipina@ehu.es`
[2]Fundació Tecnocampus, Avda. Ernest Lluch 32, 08302 Mataró, Spain'
`faundez@tecnocampus.cat`
[3]Data and Signal Processing Research Group, University of Vic – Central University of Catalonia, Sagrada Família 7, 08500 Vic, Spain
`jordi.sole@uvic.cat`
[4]Universidad de Las Palmas de Gran Canaria, IDeTIC
`jalonso@dsc.ulpgc.es`
[5]Brn GuABIAN, Association for Personal Autonomy
`guabiangipuzkoa@gmail.com`

**Abstract.** Most of medical developments require the ability to identify samples that are anomalous with respect to a target group or control group, in the sense they could belong to a new, previously unseen class or are not class data. In this case when there are not enough data to train two-class One-class classifi-cation appear like an available solution. On the other hand non-linear approaches could give very useful information. The aim of our project is to contribute to earlier diagnosis of AD and better estimates of its severity by using automatic analysis performed through new biomarkers extracted from speech signal. The methods selected in this case are speech biomarkers oriented to Spontaneous Speech and Emotional Response Analysis. In this approach One-class classifiers and two-class classifiers are analyzed. The use of information about outlier and Fractal Dimension features improves the system performance.

**Keywords:** One-class classifier, Nonlinear Speech Processing, Alzheimer disease diagnosis, Spontaneous Speech, Fractal Dimensions.

## 1 Introduction

Many applications (most of medical developments) re-quire the ability to identify samples that are anomalous with respect to a target group or control group, in the sense they belong to a new, previously unseen class or are not class data as in not common diseases or environment with very few population. In this case there are not enough data to train two-class models classifier, as in pilot studies, one possible approach to this kind of verification or identification problems is one-class classification, learning a description of the target class concerned based solely on data from this class. One-class classification problem [1] differs from multi-class classifier in one essential aspect. In one-class classification it is assumed that only information of one of the classes, the target class, is available. This means that just example objects of the target class can be used and that no information about the other class of outlier objects is present. The different terms such as fault detection, anomaly detection, novelty detection and outlier detection originate from the different applications to which one class classification can be applied. The boundary between the two classes has to be estimated from data of only control class. The task is to define a boundary around the target class, such that it accepts as much of the target/control objects as possible, while it minimizes the chance of accepting outlier objects. In the literature a large number of different terms have been used for this problem. The term one-class classification originates from [2]. The application is as follows: The first application for one class classification (also called data description as it forms the boundary around the whole available data) is outlier detection, to detect uncharacteristic objects from a dataset. Secondly, data description can be used for a classi-fication problem where one of the classes is sampled very well, while the other class is severely under sampled. The measurements on the under sampled class might be very expensive or difficult to obtain. Finally, the last possible use of the outlier detection is the comparison of two data sets. Assume that a classifier has been trained (in a long and difficult optimization process) on some (possibly expensive) data [3]. As explained above, the second application of outlier detection is in the classification problem where one of the classes is sampled very well but it is very hard and expensive, if not impossible, to obtain the data of the second class. One of the major difficulties inherent in the data (as in many medical diagnostic applications) is this highly skewed class distribution. The problem of imbalanced datasets is particularly crucial in applications where the goal is to maximize recognition of the minority class.

Alzheimer's Disease (AD) is the most common type of dementia among the elderly. It is characterized by progres-sive and irreversible cognitive deterioration with memory loss and impairments in judgment and language, together with other cognitive deficits and behavioral symptoms. The cognitive deficits and behavioral symptoms are severe enough to limit the ability of an individual to perform eve-ryday professional, social or family activities. As the dis-ease progresses, patients develop severe disability and full dependence. An early and accurate diagnosis of AD helps patients and their families to plan for the future and offers the best opportunity to treat the symptoms of the disease. According to current criteria, the diagnosis is expressed with different degrees of certainty as possible or probable AD when dementia is present and other possible causes have been ruled out. The diagnosis of definite AD requires the demonstration of the typical AD pathological changes at autopsy [4,5,6].In addition to the loss of memory, one of the major problems caused by AD is the loss of language skills. We can meet different communication deficits in the area of language, including aphasia (difficulty in speaking and understanding) and anomia (difficulty in recognizing and naming things). The specific communication problems the patient encounters depend on the stage of the disease [6,7,8].

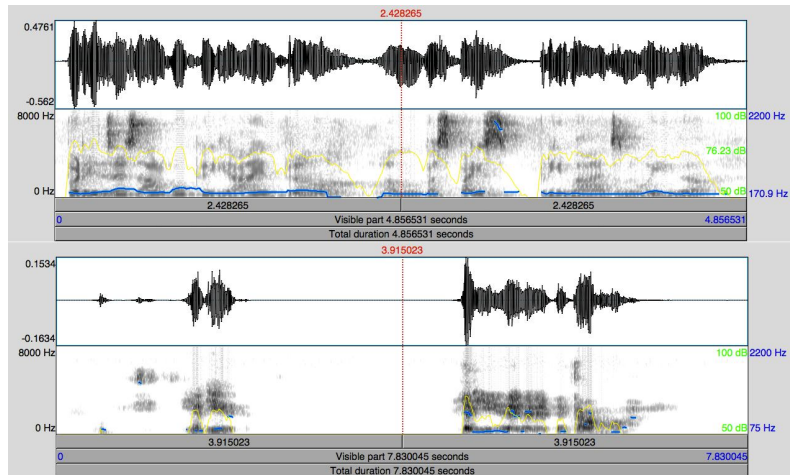

**Figure 1.** Signal and spectrogram of a control subject (top) and a subject with AD (bottom) during Spontaneous Speech (pitch in blue, intensity in yellow)

The main goal of the present work is feature search in Spontaneous Speech Analysis (ASSA) an Emotional Re-sponse Analysis (ERA) oriented to pre-clinical evaluation for the definition of test for AD diagnosis. These features will define control group (CR) and AD disease. Non-invasive Intelligent Techniques of diagnosis may become valuable tools for early detection of dementia. Moreover, these techniques are very low-cost and do not require extensive infrastructure or the availability of medical equipment. They are thus capable of yielding information easily, quick-ly, and inexpensively [9,10]. This study is focuses on early AD detection and its objective is the identification of AD in the pre-clinical (before first symptoms), Mild Cognitive Impairment (MCI) and prodromic (some very early symptoms but no dementia) stages. The research presented here is a com-plementary preliminary experiment to define thresholds for a number of biomarkers related to spontaneous speech. Feature search in this work is oriented to pre-clinical evalu-ation for the definition of test for AD diagnosis Obtained data will complement the biomarkers of each person [11]. stomach. In addition to the loss of memory, one of the major problems caused by AD is the loss of language skills (Figure 1, [12]). We can meet different communication deficits in the area of language, including aphasia (difficulty in speaking and understanding) and in Emotional Response [13-16].

## 2. Materials

This study is focused on early AD detection and its ob-jective is the identification of AD in the pre-clinical (before first symptoms) and prodromic (some very early symptoms but no dementia) stages. The research presented here is a complementary preliminary experiment to define thresholds for a number of biomarkers related to spontaneous speech. Feature search in this work is oriented to pre-clinical evaluation for the definition of test for AD diagnosis. Obtained data will complement the biomarkers of each person. Trying to develop a new methodology applicable to a wide range of individuals of different sex, age, language and cultural and social background, we have built up a multicultural and multilingual (English, French, Spanish, Catalan, Basque, Chinese, Arabian and Portuguese) database with video recordings of 50 healthy and 20 AD patients (with a prior diagnosis of Alzheimer) recorded for 12 hours and 8 hours, respectively. The age span of the individuals in the database was 20-98 years and there were 20 males and 20 females. This database is called AZTIAHO A subset of 20 AD patients was selected (68-96 years of age, 12 women, 8 men) with a distribution in the three stages of AD as follows: First Stage [ES=4], Secondary Stage [IS=10] and Tertiary stage [AS=6]. The control group (CR) was made up of 20 individuals (10 male and 10 female, aged 20-98 years) representing a wide range of speech responses. This subset of the database is called AZTIAHORE [10].

## 3. METHODS

In previous work [10,19] the goal of the experimentation was to examine the potential of the selected features to help in the automatic measurement of the degradation of Spontaneous Speech, Emotional Response and their manifestation in people with AD as compared to the control group. The approach's performance was very satisfactory and promising results for early diagnosis and classification of AD patient groups but medical doctors propose new experimentation oriented to detect mainly early stage. The goal of this new experimentation is to detect changes with regard to CR group and outliers which point to presence of AD's symptoms. One class classification will be use for this propose.

**3.1 Feature extraction**

Spoken language is one of the most important elements defining an individual's intellect, social life, and personality; it allows us to communicate with each other, share knowledge, and express our cultural and personal identity. Spoken language is the most spontaneous, natural, intuitive, and efficient method of communication among people. Therefore, the analysis by automated methods of Spontaneous Speech (SS – free and natural spoken communication), possibly combined with other methodologies, could be a useful non-invasive method for early AD diagnosis. The analysis of Spontaneous Speech fluency is based on three families of features (SSF set), obtained by the Praat software package (Praat) and software that we ourselves developed in MATLAB. For that purpose, an automatic Voice Activity Detector (VAD) has extracted voiced/unvoiced segments as parts of an acoustic signal [12,20].

### 3.1.1 Linear features

These three families of features include:
1. Duration: histogram of voiced and unvoiced segments., the average voiced/unvoiced, and variations.
2. Time domain: short time energy;
3. Frequency domain, quality: spectral centroid.

In this study, we aim to accomplish also the automatic selection of emotional speech by analyzing three families of features in speech:

1. Acoustic features: pitch, standard deviation of pitch, max and min pitch, intensity, standard deviation of intensity, max and min intensity, period mean, period standard deviation, and Root Mean Square amplitude (RMS);
2. Voice quality features: shimmer, local jitter, Noise-to-Harmonics Ratio (NHR), Harmonics-to-Noise Ratio (HNR) and autocorrelation;
3. Duration features: locally voiceless frames, degree of voice breaks.
4. Emotional Temperature [10].

### 3.1.2 Higuchi Fractal Dimension

Most of the fractal systems have a characteristic called self-similarity. An object is self-similar if a close-up examination of the object reveals that it is composed of smaller versions of itself. Self-similarity can be quantified as a relative measure of the number of basic building blocks that form a pattern, and this measure is defined as the Fractal Dimension. It should be noted that the Fractal Dimension of natural phenomena is only measurable using statistical approaches. Consequently, there exists no precise reference of the Fractal Dimension value that a given waveform should have. In addition, speech waveforms are not stationary, so most ASR techniques employ short sections of the signal in order to extract features from the waveform. This means that one plausible technique for extracting features from speech waveforms, for the purpose of recognising different phonemes, is to divide the signal in short chunks and calculate the features for each chunk. This was the approach we adopted. In other words, we calculated the Fractal Dimension of short segments of the waveform and observed the evolution of the obtained values along the whole signal, with the aim of finding in it fractal characteristics that could help in identifying different elements of the spoken message.There are several algorithms for measuring the Fractal Dimension. In this current work we focus on the alternatives which are specially suited for time series analysis and which don't need previous modelling of the system. One of these algorithms is Higuchi [21] named from his author. Higuchi and Castiglioni were chose because it has been reported to be more accurate in previous works with under-resourced conditions [10,22]. Higuchi [18] proposed an algorithm for measuring the Fractal Dimension of discrete time sequences directly from the time series *x(1),x(2),…,x(n)*. Without going into detail, the algorithm calculates the length $L_m(k)$ (see Equation 1) for each value of m and k covering all the series.

$$L_m(k) = \frac{\sum_{i=1}^{\lfloor \frac{N-m}{k} \rfloor} |x(m+ik) - x(m+(i-1)k)|(n-1)}{\lfloor \frac{N-m}{k} \rfloor k} \quad (1)$$

After that, a sum of all the lengths $L_m(k)$ for each k is determined with Equation 2.

$$L(k) = \sum_{m=1}^{k} L_m(k) \quad (2)$$

And finally, the slope of the curve ln(L(k))/ln(1/k) is estimated using least squares linear best fit, and the result is the Higuchi Fractal Dimension (HFD).

The selection of an appropriate window size to be used during the experiments is essential. Broadly speaking, the Fractal Dimension is a tool for attempting to capture the dynamics of the system. With a short window, the estimation is highly local and adapts fast to the changes in the waveform. When the window is longer, some details are lost but the Fractal Dimension better anticipates the characteristics of the signal. Additionally, previous studies that take into account the window size of similar dimension estimations [23-25] suggest that a bigger window could be useful in some cases. Consequently, four window-sizes of 160, 320 and 1280 points will be analyzed.

**3.1.3 Features sets**

In the experimentation, three families of feature sets will be used:

1. SSF+EF: Spontaneous Speech features and Emotional Speech features
2. SSF+EF+ET: SSF+EF and Emotional Temperature
3. SSF+EF+ET+VHD: SSF+EF, Emotional Temperature, Higuchi Fractal Dimension and its maximum, minimum, variance, standard deviation

**3.1.4 Automatic Classification**

The main goal of the present work is feature search in Spontaneous Speech oriented to pre-clinical evaluation for the definition of tests for AD diagnosis. These features will define CR group and the three AD levels. Moreover a secondary goal will be the optimization of computational cost oriented to real time applications. Thus automatic classification will be modelled in this sense. Two different classifier will be be evaluated:

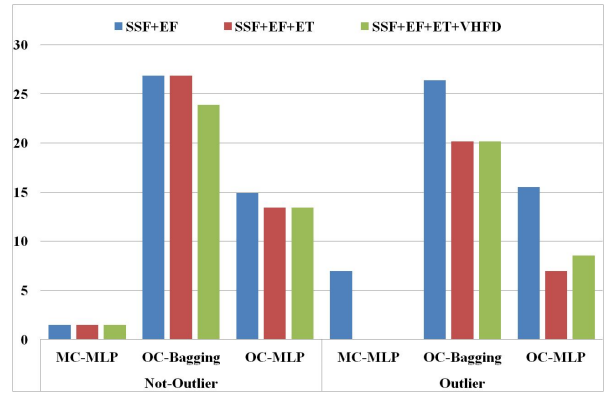

**Figure 2.** %CER global results for Multi-class and One-class classifiers

1. Multi-class classifier: Multi Layer Perceptron (MLP) with Neuron Number in Hidden Layer (NNHL) = max(Attribute/Number+Classes/Number) and Training Step (TS) NNHL*10.
2. One-Class classifier: The base classifiers to be used in the experimentation were Bagging and MLP.

WEKA [29] software was used in carrying out the experiments. The results were evaluated using Classification Error Rate (CER) and Accuracy (Acc). For the training and validation steps, we used k-fold cross-validation with k=10. Cross validation is a robust validation for variable selection [30].

## 4. Results and discussion

The task was Automatic Classification, with the classifi-cation targets being: healthy speakers without neurological pathologies and speakers diagnosed with AD. The experimentation was carried out with AZTIAHORE. The results have been analyzed with regard to the feature set described in III. Experimentation has been divided to test One-class classifier only with speech samples from CR group and with information about outliers (patients with AD) and Multi-class classifier (MLP). The results are shown in Figure 2,3. In preliminary experiments and based in previous works a window-size of 320 samples have been selected. The results are satisfactory for this study because they obtained very good results for all feature sets (Figure 2.). Most of them have about 100% of Accuracy for CR group. However it must be highlight that despite global result are optimum there is a lack for experiment without outlier. In this case results are very poor for people with AD because there are not enough samples to train properly the MLP. The best results are obtained for SSF+EF+ET+VHFD.

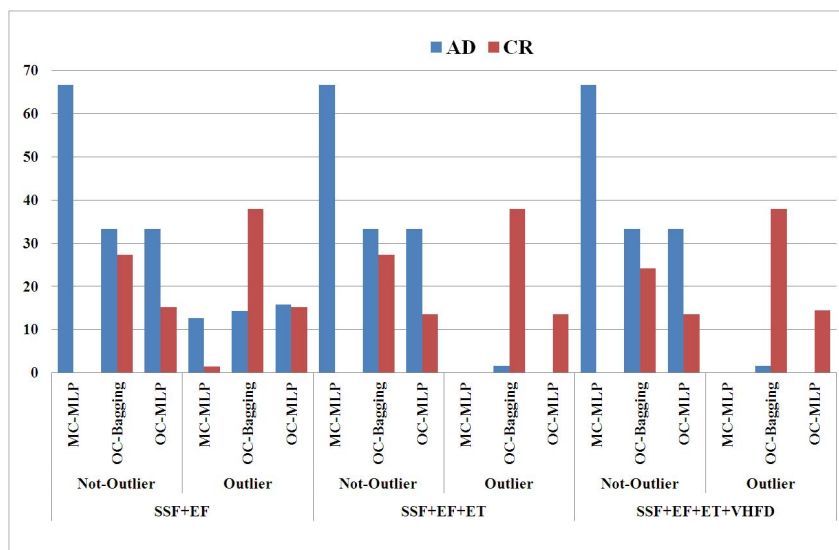

Figure 4. %CER classes' results for Multi-class and One-class classifiers

One-class classifier outperforms MLP when outliers group is not used in the training process. The results are satisfactory for this study in the case of MLP classifier. Bagging paradigm presents lower computational cost but very poor results. The new fractal features improve the system but they can improve Bagging performance. The best results are obtained for integral feature set, which mixes features relative to Spontaneous Speech and Emo-tional Response (SSF+VHFD+EF+ET). ET appears also as a powerful feature to discriminate AD. The results are satisfactory for this study because they obtained very good results not only for MLP classifier but also for Bagging, which presents lower computational cost. One-class classifier with MLP is the option which shows better performance in all cases. Moreover this model out-performs MLP when there is a lack of appropriate AD samples.

## 5. Conclusions

One-class classification has been used for detection of AD symptoms in spontaneous speech with under-resourced condition. In one-class classification it is assumed that only information of one of the classes, the target class, is available. In this work we explore the problem of imbalanced datasets by using information about outlier and Fractal Dimension features improves the system performance and Emotional Temperature. One-class classifier with MLP outperforms two-class classifier when there are very few AD samples. The approach of this work complements the previous multi-class modelling (two class) and is robust in terms of capturing the dynamics of the speech, and it offers many advantages in terms of be easier to compare the power of the new features against the previous ones. In future works we will introduce new features relatives to speech modelling oriented to standard medical tests for AD diagnosis and to emotion response analysis.

## Acknowledgement

This work has been supported by FEDER and Ministerio de ciencia e Innovación, TEC2012-38630-C04-03 and the University of the Basque Country by EHU-14/58. The described research was performed in laboratories supported by the SIX project; the registration number CZ.1.05/2.1.00/03.0072, the operational program Research and Development for Innovation.